\documentclass[11pt]{article}
\usepackage[left=80pt, right=80pt, top=90pt]{geometry}
\usepackage{amsmath}
\usepackage[multiple]{footmisc}
\usepackage{hyperref}
\usepackage{titling}
\makeatletter
\newcommand{\raisemath}[1]{\mathpalette{\raisem@th{#1}}}
\newcommand{\raisem@th}[3]{\raisebox{#1}{$#2#3$}}
\makeatother
	
\title{\bf Investigation of the solution of a system of partial differential equations with periodic coefficients}
\author{by A. Potier \\ (Translated from the French, edited and annotated by Edward F. Kuester)}
\date{}

\begin{document}
	
\begin{titlingpage}
	\maketitle
\begin{abstract}
	This is an English translation of a paper by the French physicist Alfred Potier (1840-1905) that originally appeared 150 years ago [A. Potier, ``Recherches sur l'int\'egration d'un syst\`eme d'\'equations aux diff\'erentielles partielles \`a coefficients p\'eriodiques,'' Comptes Rendus de l'Association Fran\c{c}aise pour l'Avancement des Sciences (Bordeaux), vol. 1, pp. 255-272 (1872)]. The paper presents an analysis of wave propagation through a periodic medium by a method that in many ways anticipated the technique of multiple-scale homogenization by more than a century.
\end{abstract}	
\end{titlingpage}
	
\section*{\centering Translator's Preface}

\renewcommand{\thefootnote}{\alph{footnote}}

This is an English translation of a paper by the French physicist Alfred Potier (1840-1905) that originally appeared in 1872\footnote{A. Potier, ``Recherches sur l'int\'egration d'un syst\`eme d'\'equations aux diff\'erentielles partielles \`a coefficients p\'eriodiques,'' \emph{Comptes Rendus de l'Association Fran\c{c}aise pour l'Avancement des Sciences (Bordeaux)}, vol. 1, pp. 255-272 (1872): \href{https://gallica.bnf.fr/ark:/12148/bpt6k201148c/f258.item}{https://gallica.bnf.fr/ark:/12148/bpt6k201148c/f258.item} .} and was reprinted in a collection of his works forty years later.\footnote{A. Potier, \emph{M\'emoires sur l'\'Electricit\'e et l'Optique}. Paris: Gauthier-Villars, 1912, pp. 239-256.} The paper presents an analysis of wave propagation through a periodic medium by a method that in many ways anticipated the technique of multiple-scale homogenization by more than a century.\footnote{A. Bensoussan, J.-L. Lions and G. Papanicolaou, \emph{Asymptotic Analysis for Periodic Structures}.  Amsterdam: North-Holland, 1978.}\footnote{E. Sanchez-Palencia, \emph{Non-Homogeneous Media and Vibration Theory}. Berlin: Springer-Verlag, 1980.}
Indeed, this paper appeared nearly a decade before Floquet's work on ordinary differential equations with periodic coefficients,\footnote{G. Floquet, ``Sur les \'equations diff\'erentielles lin\'eaires \`a coefficients p\'eriodiques,'' \emph{Comptes Rendus Acad. Sci. (Paris)}, vol. 91, pp. 880-882 (1880).}\footnote{G. Floquet, ``Sur les \'equations diff\'erentielles lin\'eaires \`a coefficients p\'eriodiques,'' \emph{Ann. Sci. \'Ecole Norm. Sup.}, ser. 2, vol. 12, pp. 47-88 (1883).}\footnote{G. Floquet, ``Sur une classe d'\'equations diff\'erentielles lin\'eaires non homog\`enes,'' \emph{Ann. Sci. \'Ecole Norm. Sup.}, ser. 3, vol. 4, pp. 111-128 (1887).} and it would be more than 50 years before the appearance of Bloch's paper treating the problem of Schr\"odinger's equation in a three-dimensionally periodic potential.\footnote{F. Bloch, ``\"Uber die Quantenmechanik der Elektronen in Kristallgittern,'' \emph{Zeits. Physik}, vol. 52, pp. 555-600 (1928).}

Neither of these authors seems to have been aware of Potier's work, and his paper remained essentially unknown until almost the end of the twentieth century. Although Poincar\'e made passing reference to it in his lectures on the mathematical theory of light,\footnote{H. Poincar\'e, \emph{Le\c{c}ons sur la Th\'eorie Math\'ematique de la Lumi\`ere}. Paris: Georges Carr\'e, 1889, pp. 258 and 318.} the only other citation to it before 1994 that I have found is in a paper by Chipart in 1924.\footnote{H. Chipart, ``Sur la propagation de la lumi\`ere dans les milieux \`a structure p\'eriodique,'' \emph{Comptes Rendus Acad. Sci. (Paris)}, vol. 178, pp. 319-321 (1924).}
This lack of citations is perhaps not too surprising, since Potier's paper was published in a fairly obscure journal. The paper also suffers from a significant number of typographical errors and confusing notations that can hamper understanding of it, but I believe that it merits a wider readership, and so present this translation.

The connection of Potier's work to the modern theory of multiple-scale homogenization is perhaps not quite a direct one. He assumes a quasi-plane wave to be propagating in the medium that takes the form of what we now call a Floquet-Bloch wave---an exponential function multiplied by a function periodic with the same periodicity as the constitutive properties of the medium. In modern homogenization technique, that exponential function would be replaced by a more arbitrary function that varies slowly with respect to the period of the medium. But what ties Potier's analysis to the modern theory (and what distinguishes his construction of the solution from that of Floquet or Bloch) is his expansion of the periodic function in a series of ascending powers of frequency, whose coefficients he proceeds to solve for in a recursive manner. At each stage of the process, the solution of a static boundary value problem is required to determine these coefficients, in the same way as is done today.$^{\rm c,d}$ Although the periodicity of the medium does not appear explicitly in his expansion, it does so in an implicit way that he did not explore in detail.

In 1872, the publication of Maxwell's theory of electromagnetic fields had taken place only seven years earlier, and the theory had yet to achieve wide acceptance or receive full experimental confirmation. Thus, Potier regarded light as an elastic wave phenomenon propagating in the ether. The modern reader must allow for that, and can think of the vector whose cartesian components are $(u,v,w)$ as the electric field $\mathbf{E}$. Of course, the equations of elasticity are not quite the same as those describing the electromagnetic field in an inhomogeneous medium, but the method laid out in this paper could easily be adapted to treat Maxwell's equations as well.

For clarity and easier readability, I have made the mathematical notation somewhat more consistent with that normally used today. The ordinary derivative notations $d/dx$, $d/dy$ and $d/dz$ have been replaced by the corresponding partial derivatives $\partial/\partial x$, $\partial/\partial y$ and $\partial/\partial z$. There are also, in the French original, some infelicities of notation wherein the same symbol is used to denote different things in different parts of the paper. I have introduced new notations where needed to avoid confusion arising from that, or simply to conform to common usage today. I have endeavored to be as faithful to the original French as I could, but in a few places I have taken liberties in the translation in the interest of clarity. Translator's notes were inserted when it seemed to me that more extensive explanation was needed. I have inserted equation numbers (T1), (T2), etc. when I needed to refer to an equation that Potier left unnumbered. All equations in the paper have been reviewed; corrections of the typographical errors I found in them were made where necessary, and such equations marked with a $(*)$.

I will be happy to be informed of any incorrect translations or further errors that I have missed in the original paper, and will include them in updated versions of this document.

\begin{flushright}
	Edward F. Kuester \\
	University of Colorado Boulder \\
	\href{mailto:edward.kuester@colorado.edu}{edward.kuester@colorado.edu} \\
	\today
\end{flushright}

\clearpage

\renewcommand{\thefootnote}{\arabic{footnote}}
\setcounter{footnote}{0}

\begin{center}
	{\Large\bf Investigation of the solution of a system of partial differential equations with periodic coefficients}\protect\footnote{Session of 12 September 1872.} \\
\vspace{0.2in}
{\large A. Potier} \\
(Mining Engineer)

\vspace{0.2in}
\end{center}

The various hypotheses that can be advanced about the nature of the medium that transmits light ultimately rest on the differential equations governing the movement of the molecules of this medium, and whose solutions can be subjected to experimental verification.

This in turn will allow us to accept or reject the proposed differential equations, with the proviso, however, that the method used to solve them is valid without question. Now among these equations, those whose coefficients are periodic have recently prompted a discussion concerning the method of solution; I will briefly indicate the crux of this debate, and then I will propose a different method that appears to me to be beyond dispute.

In a memoir published in the \emph{Journal de Math\'{e}matiques Pures et Appliqu\'{e}es},\footnote{
E. Sarrau, ``Sur la propagation et la polarisation de la lumi\`{e}re dans les cristaux,'' {\em J. Math. Pures Appl.}, ser. 2, vol.~12, pp. 1-46 (1867) and vol.~13, pp. 59-110 (1868).} M. Sarrau, having given reasons to believe that the coefficients in the differential equations for the movement of the ether in crystalline media must be periodic, sought to obtain (using a method given by Cauchy) the differential equations with constant coefficients that must be satisfied by the average values of the displacements. This method consists in expanding the given periodic coefficients, as well as the components of the molecular displacement of the ether, in series of imaginary exponentials. The equations thereby obtained are then separated into as many equations as the number of terms it is necessary to keep in these series, that is, into an infinite number of equations (admittedly linear in the unknowns); it is easy to see that even though the series corresponding to the given data (that is, the series into which the periodic coefficients characterizing the medium are expanded) have only a finite number of terms, we nevertheless have to take the series for the displacements to be infinite, so that the number of equations of first order to be solved can never be finite; it is therefore not certain that we can extend to this infinite set of equations the conclusions that can be drawn from a study of a finite number of equations.

When we limit ourselves to the practical case when the periodic coefficients reduce to a single one, all these difficulties remain, and M. de Saint-Venant in his memoir ``On various ways of presenting the theory of light waves''\footnote{A. J. C. Barr\'e de Saint-Venant, ``Sur les divers mani\`eres de pr\'esenter la th\'eorie des ondes lumineuses,'' \emph{Ann. Chim. Phys.}, ser. 4, vol. 25, pp. 335-381 (1872).} even seems to think that the method is not exact. Indeed, here is how the eminent mathematician puts it:

\begin{quotation}
``\ldots the only periodic coefficient that enters into his [M. Sarrau's] equations is the inverse $\frac{1}{\rho}$ of the density of the ether or more generally, as he assumes, a function of $\rho$ called $e$. It is this function of the density appearing in the right side of these equations that he replaces by its constant average value plus periodic terms, making similar replacements for the displacements $u$, $v$, $w$, from which he obtains by Cauchy's method (and for various crystalline forms) the equations for the average displacements. They yield laws of proportionality for double refraction, rotatory polarization, etc. that agree with experiment.

``But the factor $\rho$ could just as well have stayed on the left side of the equation. When, according to the hypothesis, we equate this variable density to a periodic function, instead of its inverse $\frac{1}{\rho}$, since $\rho$ multiplies the second derivative with respect to time while on the other hand $\frac{1}{\rho}$ multiplies the second derivative with respect to the spatial coordinates, we might imagine that the result would not be the same and that it would yield a law proportional to something other than what was sought. This is in fact what happens. The second method (the more natural of the two) leads to the wrong behavior, contrary to experiment; namely, that the coefficient of birefringence, instead of being essentially independent of wavelength, would vary as its inverse square, whereas the rotatory power that varies as the inverse square of wavelength would vary as its inverse fourth power.''
\end{quotation}

In short, Cauchy's method of solution applied to these same equations would lead to two different results depending on whether a factor is placed on one side of the equation or the other; if this was really the case, we would be justified in saying that equations with periodic coefficients cannot be solved, and as a consequence cannot be tested experimentally.

Therefore, even though this question is already an old one, I hope the following work in which I will propose another method will be of some interest, not least because I have placed the factor $\rho$ precisely with the derivatives with respect to time, but nevertheless the resulting solution is in agreement with known laws.

\vspace{0.2in}

In what follows, $u$, $v$ and $w$ denote the components of displacement of a molecule of the ether. The coefficient of elastic density $\rho$ is assumed to vary periodically, so it is consequently sufficient to know its value in the interior of a certain parallelepiped in order to deduce its value over all space. This parallelepiped will be called the elementary volume. Moreover, since we are only concerned with periodic motion in the theory of light, we can always put
\[
\frac{d^2 u}{dt^2} = - \omega^2 u , \qquad \frac{d^2 v}{dt^2} = - \omega^2 v , \qquad \frac{d^2 w}{dt^2} = - \omega^2 w
\]
where we have denoted the quotient $\displaystyle\frac{2\pi}{\tau}$ by $\omega$, $\tau$ being the period of the oscillations.\footnote{$\omega$ was denoted $\alpha$ in the original (\emph{Translator's note}).}

We want to find out whether the partial differential equations
\[
\nabla^2 u - \frac{\partial \theta}{\partial x} = - \rho \omega^2 u ,
\]
\[
\nabla^2 v - \frac{\partial \theta}{\partial y} = - \rho \omega^2 v ,
\]
\[
\nabla^2 w - \frac{\partial \theta}{\partial z} = - \rho \omega^2 w , \eqno{({\rm T1})}
\]
in which the symbol $\nabla^2$ stands for
\[
\frac{\partial^2}{\partial x^2} + \frac{\partial^2}{\partial y^2} + \frac{\partial^2}{\partial z^2}
\]
and
\[
\theta = \frac{\partial u}{\partial x} + \frac{\partial v}{\partial y} + \frac{\partial w}{\partial z} ,
\]
possess a system of solutions of the form
\[
u = {\rm A} e^{mx + ny +pz} , \qquad v = {\rm B} e^{mx + ny +pz} , \qquad w = {\rm C} e^{mx + ny +pz} \eqno{(\rm T2)}
\]
with $m$, $n$ and $p$ being constants, while ${\rm A}$, ${\rm B}$ and ${\rm C}$ are periodic functions with the same periodicities as $\rho$.

We see immediately that if $m$, $n$, $p$ differ from 0, they must depend on $\omega$, and the same is true for ${\rm A}$, ${\rm B}$ and ${\rm C}$. In addition, the equations above keep the same form when the coordinate axes are changed; things can always be arranged so that $m$ and $n$ are zero.\footnote{It seems to me that this cannot be true for a general anisotropic periodic medium. Potier appears to assume that the medium is periodic along each of the cartesian coordinates $x$, $y$ and $z$. If the variation of material properties is different in these three directions, and nonseparable, then the case $m=n=0$ can only give behavior of the solutions that is typical, but not universal (\emph{Translator's note}).} Thus, we only need to calculate ${\rm A}$, ${\rm B}$, ${\rm C}$ and $p$; substituting ${\rm A} e^{pz}$, ${\rm B} e^{pz}$ and ${\rm C} e^{pz}$ for $u$, $v$ and $w$ in the equations under consideration, they become
\begin{eqnarray*}
\nabla^2 {\rm A} - \frac{\partial \Theta}{\partial x} + p \left( 2 \frac{\partial {\rm A}}{\partial z} - \frac{\partial {\rm C}}{\partial x} \right) + \left( p^2 + \rho \omega^2 \right) {\rm A} & = & 0 \\
\nabla^2 {\rm B} - \frac{\partial \Theta}{\partial y} + p \left( 2 \frac{\partial {\rm B}}{\partial z} - \frac{\partial {\rm C}}{\partial y} \right) + \left( p^2 + \rho \omega^2 \right) {\rm B} & = & 0 \\
\nabla^2 {\rm C} - \frac{\partial \Theta}{\partial z} - p \left( \frac{\partial {\rm A}}{\partial x} + \frac{\partial {\rm B}}{\partial y} \right) + \rho \omega^2 {\rm C} & = & 0
\end{eqnarray*}
with
\[
\Theta = \frac{\partial {\rm A}}{\partial x} + \frac{\partial {\rm B}}{\partial y} + \frac{\partial {\rm C}}{\partial z}
\]

We expand ${\rm A}$, ${\rm B}$, ${\rm C}$, $p$ and $p^2$ in powers of $\omega$ by putting\footnote{Evidently, ${\rm P}_2 = p_1^2$, ${\rm P}_3 = 2p_1p_2$, ${\rm P}_4 = p_2^2 + 2p_1 p_3$, etc. (\emph{Translator's note}).}
\begin{eqnarray*}
{\rm A} & = & {\rm A}_0 + {\rm A}_1 \omega + {\rm A}_2 \omega^2 + \ldots \\
{\rm B} & = & {\rm B}_0 + B_1 \omega + {\rm B}_2 \omega^2 + \ldots \\
{\rm C} & = & {\rm C}_0 + {\rm C}_1 \omega + {\rm C}_2 \omega^2 + \ldots \\
p & = & p_1 \omega + p_2 \omega^2 + \ldots \\
p^2 & = & {\rm P}_2 \omega^2 + {\rm P}_3 \omega^3 + \ldots
\end{eqnarray*}
If the proposed form of solution is acceptable, we should be able to determine ${\rm A}$, ${\rm B}$ and ${\rm C}$ as functions periodic like $\rho$.

Substitution of these expressions into the differential equations\footnote{
	and equating the coefficients of each power of $\omega$ separately (\emph{Translator's note})} gives
\begin{eqnarray*}
\nabla^2 {\rm A}_0 - \frac{\partial \Theta_0}{\partial x} = 0, & \quad & \nabla^2 {\rm A}_1 - \frac{\partial \Theta_1}{\partial x} + p_1 \left( 2 \frac{\partial {\rm A}_0}{\partial z} - \frac{\partial {\rm C}_0}{\partial x} \right) = 0 \\
\nabla^2 {\rm B}_0 - \frac{\partial \Theta_0}{\partial y} = 0, & \quad & \nabla^2 {\rm B}_1 - \frac{\partial \Theta_1}{\partial y} + p_1 \left( 2 \frac{\partial {\rm B}_0}{\partial z} - \frac{\partial {\rm C}_0}{\partial y} \right) = 0 \\
\nabla^2 {\rm C}_0 - \frac{\partial \Theta_0}{\partial z} = 0, & \quad & \nabla^2 {\rm C}_1 - \frac{\partial \Theta_1}{\partial z} - p_1 \left( \frac{\partial {\rm A}_0}{\partial x} + \frac{\partial {\rm B}_0}{\partial y} \right) = 0
\end{eqnarray*}
and in general\footnote{I.~e., for $n \geq 2$. Potier evidently intends that $\displaystyle
	\Theta_n = \frac{\partial {\rm A}_n}{\partial x} + \frac{\partial {\rm B}_n}{\partial y} + \frac{\partial {\rm C}_n}{\partial z}
	$. Note that the subscript $n$ used here is not the same as the quantity $n$ that appears in equation (T2) (\emph{Translator's note}).}
\begin{equation}
\left.
\begin{array}{ccc}
\nabla^2 {\rm A}_n - \displaystyle \frac{\partial \Theta_n}{\partial x} + \displaystyle\sum_{i=0}^{n-1} \left( 2 \frac{\partial {\rm A}_i}{\partial z} - \frac{\partial {\rm C}_i}{\partial x} \right) p_{n-i} + \displaystyle\sum_{j=0}^{n-2} {\rm P}_{n-j} {\rm A}_j + \rho {\rm A}_{n-2} & = & 0  \\
\nabla^2 {\rm B}_n - \displaystyle \frac{\partial \Theta_n}{\partial y} + \displaystyle\sum_{i=0}^{n-1} \left( 2 \frac{\partial {\rm B}_i}{\partial z} - \frac{\partial {\rm C}_i}{\partial y} \right) p_{n-i} + \displaystyle\sum_{j=0}^{n-2} {\rm P}_{n-j} {\rm B}_j + \rho {\rm B}_{n-2} & = & 0 \\
\nabla^2 {\rm C}_n - \displaystyle \frac{\partial \Theta_n}{\partial z} - \displaystyle\sum_{i=0}^{n-1} \left( \frac{\partial {\rm A}_i}{\partial x} + \frac{\partial {\rm B}_i}{\partial y} \right) p_{n-i} + \rho {\rm C}_{n-2} & = & 0 
\end{array}
\right\} \label{e1}
\end{equation}

We can thus iteratively determine ${\rm A}$, ${\rm B}$ and ${\rm C}$ to a better and better approximation if the equations above are compatible. The compatibility condition will be obtained by differentiating the first with respect to $x$, the second with respect to $y$, the third with respect to $z$ and adding. The resulting sum will contain the various groups of terms $\left( \nabla^2 {\rm A}_n - \displaystyle\frac{\partial \Theta_n}{\partial x} \right) \ldots$ that should be replaced by expressions for them that follow from (1); what results is the equation
\begin{eqnarray}
\lefteqn{\frac{\partial}{\partial x} \left( \rho {\rm A}_{n-2} \right) + \frac{\partial}{\partial y} \left( \rho {\rm B}_{n-2} \right) + \frac{\partial}{\partial z} \left( \rho {\rm C}_{n-2} \right)} \nonumber \\
&& + \rho \left( p_1 {\rm C}_{n-3} + p_2 {\rm C}_{n-4} + \ldots + p_{n-2} {\rm C}_0 \right) = 0 \label{e2}
\end{eqnarray}
By the way, this equation can be deduced directly from the equation\footnote{This equation follows from equation (T1) above, taking the $x$, $y$ and $z$ derivative of each line respectively, and adding them up (\emph{Translator's note}).}
\[
\frac{\partial (\rho u)}{\partial x} + \frac{\partial (\rho v)}{\partial y} + \frac{\partial (\rho w)}{\partial z} = 0 \eqno{(*)}
\]

Next, we impose the condition that the functions ${\rm A}$, ${\rm B}$ and ${\rm C}$ be periodic. Because of the fact that any integral taken over the elementary volume of a derivative of a periodic function will be zero, the integral over this volume of the non-differentiated parts of equations (\ref{e1}) and (\ref{e2}) must be zero. We will adopt the convention that the notation $\left[ \Phi \right]$ represents the average value of a function $\Phi$ taken over the elementary volume:
\[
\left[ \Phi \right] = \frac{1}{V_p} \int\int\int \Phi \, dx \, dy \, dz \equiv \frac{1}{V_p} \int \Phi \, d\varpi \eqno{(*)}
\]
where $V_p = \int \, d\varpi$ is the volume of the elementary parallelepiped.
We then have\footnote{For $n \geq 2$; if $n=0$ or 1, the summations should be omitted (\emph{Translator's note}).}
\begin{equation}
\left[ \sum_{j=0}^{n-2} {\rm P}_{n-j} {\rm A}_{j} + \rho {\rm A}_{n-2} \right] = 0, \quad \left[ \sum_{j=0}^{n-2} {\rm P}_
{n-j} {\rm B}_{j} + \rho {\rm B}_{n-2} \right] = 0, \quad \left[ \rho {\rm C}_{n-2} \right] = 0 . \ (*)
\label{e3}
\end{equation}
When conditions (\ref{e2}) and (\ref{e3}) are satisfied, the system (\ref{e1}) will have periodic solutions.\footnote{It appears to me that while conditions (2) and (3) are necessary, Potier has not proved them to be sufficient for periodic solutions to exist (\emph{Translator's note}).}

The first equations to be solved are for ${\rm A}_0$, ${\rm B}_0$ and ${\rm C}_0$, which must satisfy the conditions
\[
\nabla^2 {\rm A}_0 - \frac{\partial \Theta_0}{\partial x} = \nabla^2 {\rm B}_0 - \frac{\partial \Theta_0}{\partial y} = \nabla^2 {\rm C}_0 - \frac{\partial \Theta_0}{\partial z} = 0
\]
\[
\frac{\partial}{\partial x} \left( \rho {\rm A}_{0} \right) + \frac{\partial}{\partial y} \left( \rho {\rm B}_{0} \right) + \frac{\partial}{\partial z} \left( \rho {\rm C}_{0} \right) = 0
\eqno{({\rm T3})}
\]
\[
\left[ ({\rm P}_2 + \rho) {\rm A}_0 \right] = \left[ ({\rm P}_2 + \rho) {\rm B}_0 \right] = \left[ \rho {\rm C}_0 \right] = 0
\]
From the first three equations we find that
\[
{\rm A}_0 = a_0 + \frac{\partial \varphi}{\partial x} , \quad {\rm B}_0 = b_0 + \frac{\partial \varphi}{\partial y} , \quad {\rm C}_0 = c_0 + \frac{\partial \varphi}{\partial z}
\]
where $\varphi$ is a periodic function\footnote{The functions $\varphi$, $\psi$ and $\chi$ called periodic in this paper are also assumed to have an average values of 0 over the elementary volume; see Appendix 1. Since the integral of the derivative of a periodic function over the elementary volume is zero, the quantities $a_0$, $b_0$ and $c_0$ represent the average values of ${\rm A}_0$, ${\rm B}_0$ and ${\rm C}_0$ respectively (\emph{Translator's note}).} and $a_0$, $b_0$ and $c_0$ are constants, satisfying the equations
\begin{equation}
\frac{\partial}{\partial x} \left( \rho \frac{\partial \varphi}{\partial x} \right) + \frac{\partial}{\partial y} \left( \rho \frac{\partial \varphi}{\partial y} \right) + \frac{\partial}{\partial z} \left( \rho \frac{\partial \varphi}{\partial z} \right) + a_0 \frac{\partial \rho}{\partial x} + b_0 \frac{\partial \rho}{\partial y} + c_0 \frac{\partial \rho}{\partial z} = 0
\label{e4}
\end{equation}
\begin{equation}
{\rm P}_2 a_0 + \left[ \rho \left( a_0 + \frac{\partial \varphi}{\partial x} \right) \right] = 0 , \ {\rm P}_2 b_0 + \left[ \rho \left( b_0 + \frac{\partial \varphi}{\partial y} \right) \right] = 0 , \ \left[ \rho \left( c_0 + \frac{\partial \varphi}{\partial z} \right) \right] = 0
\label{e5}
\end{equation}
These equations completely determine the function $\varphi$, the value of ${\rm P}_2$ and the ratios $a_0 : b_0 : c_0$ (see Appendix 1).\footnote{In fact, the function $\varphi$ as well as the constants $a_0$, $b_0$ and $c_0$ are determined only to within a common multiplying factor, since the equations they satisfy are homogeneous (\emph{Translator's note}).}

Substituting the expressions thus obtained into the equations that determine ${\rm A}_1$, ${\rm B}_1$ and ${\rm C}_1$, we have
\[
\nabla^2 {\rm A}_1 - \frac{\partial \Theta_1}{\partial x} + p_1 \frac{\partial^2 \varphi}{\partial x \partial z} = 0, \qquad \nabla^2 {\rm B}_1 - \frac{\partial \Theta_1}{\partial y} + p_1 \frac{\partial^2 \varphi}{\partial y \partial z} = 0,
\]
\[
\nabla^2 {\rm C}_1 - \frac{\partial \Theta_1}{\partial z} - p_1 \left( \frac{\partial^2 \varphi}{\partial x^2} + \frac{\partial^2 \varphi}{\partial y^2} \right) = 0 ,
\]
\[
\rho p_1 {\rm C}_0 + \frac{\partial}{\partial x} \left(  \rho {\rm A}_1 \right) + \frac{\partial}{\partial y} \left(  \rho {\rm B}_1 \right)  + \frac{\partial}{\partial z} \left(  \rho {\rm C}_1 \right)  = 0 ,
\]
\[
{\rm P}_3 a_0 + \left[ \left( {\rm P}_2 + \rho \right) {\rm A}_1 \right] = 0, \quad {\rm P}_3 b_0 + \left[ \left( {\rm P}_2 + \rho \right) {\rm B}_1 \right] = 0, \quad \left[ \rho {\rm C}_1 \right] = 0 .
\]

The solutions of the equations in the first two lines are
\begin{equation}
{\rm A}_1 = a_1 + \frac{\partial \psi}{\partial x}, \quad {\rm B}_1 = b_1 + \frac{\partial \psi}{\partial y} , \quad {\rm C}_1 = c_1 + p_1 \varphi +  \frac{\partial \psi}{\partial z}
\label{e6}
\end{equation}
in which the periodic function $\psi$ and the constants $a_1$, $b_1$ and $c_1$ are determined by the conditions
\[
a_1 \frac{\partial \rho}{\partial x} + b_1 \frac{\partial \rho}{\partial y} + c_1 \frac{\partial \rho}{\partial z} + \frac{\partial}{\partial x} \left( \rho \frac{\partial \psi}{\partial x} \right) + \frac{\partial}{\partial y} \left( \rho \frac{\partial \psi}{\partial y} \right) + \frac{\partial}{\partial z} \left( \rho \frac{\partial \psi}{\partial z} \right)
+ p_1 \frac{\partial (\rho \varphi)}{\partial z} + p_1 \rho \left( c_0 + \frac{\partial \varphi}{\partial z} \right) = 0 \quad (*)
\eqno{(\rm T4)}
\]
\[
\left.
\begin{array}{ccc}
a_1 [\rho] + {\rm P}_3 a_0 + {\rm P}_2 a_1 + \left[ \rho \displaystyle\frac{\partial \psi}{\partial x} \right] & = & 0
\quad (*)
\\
\\
b_1 [\rho] + {\rm P}_3 b_0 + {\rm P}_2 b_1 + \left[ \rho \displaystyle\frac{\partial \psi}{\partial y} \right] & = & 0
\quad (*)
\\
\\
\left[ \left( \rho c_1 + p_1 \rho \varphi \right) + \rho \displaystyle\frac{\partial \psi}{\partial z} \right] & = & 0 \qquad
\end{array}
\right\}
\eqno{(\rm T5)}
\]
(see Appendix 2).

Substituting the expressions for ${\rm A}_0$, ${\rm B}_0$, ${\rm C}_0$ and ${\rm A}_1$, ${\rm B}_1$, ${\rm C}_1$ into the equations that determine ${\rm A}_2$, ${\rm B}_2$, ${\rm C}_2$, ${\rm P}_4$, \ldots (the first of the most general form of these conditions), we arrive at:
\[
\left. \begin{array}{ccc}
\nabla^2 {\rm A}_2 - \displaystyle\frac{\partial \Theta_2}{\partial x} + \frac{\partial^2}{\partial x \partial z} \left( p_1 \psi + p_2 \varphi \right) + {\rm P}_2 a_0 + \rho {\rm A}_0 & = & 0 \qquad
\\
\\
\nabla^2 {\rm B}_2 - \displaystyle\frac{\partial \Theta_2}{\partial y} + \frac{\partial^2}{\partial y \partial z} \left( p_1 \psi + p_2 \varphi \right) + {\rm P}_2 b_0 + \rho {\rm B}_0 & = & 0 \qquad
\\
\\
\nabla^2 {\rm C}_2 - \displaystyle\frac{\partial \Theta_2}{\partial z}- \left( \frac{\partial^2}{\partial x^2} + \frac{\partial^2}{\partial y^2} \right) \left( p_1 \psi + p_2 \varphi \right) + \rho {\rm C}_0 & = & 0 . \quad (*)
\end{array}
\right\}
\eqno{({\rm T6})}
\]

These equations admit periodic solutions because of how $\varphi$ and $\psi$ were determined. We will begin by finding three periodic functions\footnote{Observe the difference here and afterwards between the italicized $A_2$, $B_2$, $C_2$ and their non-italicized versions ${\rm A}_2$, ${\rm B}_2$, ${\rm C}_2$. The constants $a_2$, $b_2$ and $c_2$ introduced below allow us to require that $[A_2] = [B_2] = [C_2] = 0$ (\emph{Translator's note}).} $A_2$, $B_2$ and $C_2$ that satisfy the equations
\[
\nabla^2 A_2 + \left( {\rm P}_2 a_0 + \rho {\rm A}_0 \right) = 0, \quad 
\nabla^2 B_2 + \left( {\rm P}_2 b_0 + \rho {\rm B}_0 \right) = 0, \quad \nabla^2 C_2 + \rho {\rm C}_0 = 0
\eqno{({\rm T7})}
\]
which, because of equations (\ref{e5}), is always uniquely possible. We then put\footnote{There is one step needed in the demonstration that the following functions satisfy equations (T6) which is perhaps not obvious. Define
	\[
	T_2 = \frac{\partial A_2}{\partial x} + \frac{\partial B_2}{\partial y} + \frac{\partial C_2}{\partial z}
	\]
	Note that this is not the same as $\Theta_2$. Now take $\partial/\partial x$ of the first part of (T7), $\partial/\partial y$ of the second, $\partial/\partial z$ of the third and add the results. By (T3), this gives $\nabla^2 T_2 = 0$. Since $T_2$ is periodic, from property 1 deduced in Appendix 1, we have that $T_2 = 0$ (\emph{Translator's note}).}
\[
{\rm A}_2 = A_2 + a_2 + \frac{\partial \chi}{\partial x} ,
\]
\[
{\rm B}_2 = B_2 + b_2 + \frac{\partial \chi}{\partial y} ,
\]
\[
{\rm C}_2 = C_2 + c_2 + \frac{\partial \chi}{\partial z} + p_1 \psi + p_2 \varphi ,
\]
and we have the following equations, deduced from the groups (\ref{e2}) and (\ref{e3}), to determine $\chi$, $a_2$, $b_2$ and $c_2$:
\[
\frac{\partial}{\partial x} \left( \rho A_2 \right) + \frac{\partial}{\partial y} \left( \rho B_2 \right) + \frac{\partial}{\partial z} \left( \rho C_2 \right) + a_2 \frac{\partial \rho}{\partial x} + b_2 \frac{\partial \rho}{\partial y} + c_2 \frac{\partial \rho}{\partial z}
\]
\[
+ \frac{\partial}{\partial x} \left( \rho \frac{\partial \chi}{\partial x} \right) + \frac{\partial}{\partial y} \left( \rho \frac{\partial \chi}{\partial y} \right) + \frac{\partial}{\partial z} \left( \rho \frac{\partial \chi}{\partial z} \right) + \frac{\partial}{\partial z} \left( p_1 \rho \psi + p_2 \rho \varphi \right) + \rho p_1 {\rm C}_1 + \rho p_2 {\rm C}_0 = 0
\]
\[
{\rm P}_4 a_0 + {\rm P}_3 a_1 + {\rm P}_2 a_2 + \left[ \rho {\rm A}_2 \right] = 0
\]
\[
{\rm P}_4 b_0 + {\rm P}_3 b_1 + {\rm P}_2 b_2 + \left[ \rho {\rm B}_2 \right] = 0
\]
\[
\left[ \rho {\rm C}_2 \right] = 0
\]
These equations, in the same way as the preceding ones, will determine the constants ${\rm P}_4$, $a_2$, $b_2$, $c_2$ and the function $\chi$. This process can be continued indefinitely.

Since the values of ${\rm P}_2$ are both negative,\footnote{See Appendix 1 (\emph{Translator's note}).} we deduce from each of them an imaginary value\footnote{at least for sufficiently small $\omega$ (\emph{Translator's note}).} for $p$ of the form $i{\rm S}$, where $i = \sqrt{-1}$, and carrying out all the operations indicated above, we find that the quantities
\[
\left.
\begin{array}{c}
a_0, b_0, c_0, \quad a_2, b_2, c_2, \quad {\rm P}_2, \varphi, {\rm P}_4, \\
 {\rm A}_0, {\rm B}_0, {\rm C}_0, \quad {\rm A}_2, {\rm B}_2, {\rm C}_2 \\
\mbox{\rm are real.}
\end{array}
\right|
\begin{array}{c}
a_1, b_1, c_1, \quad \psi, p_2, {\rm P}_3, \\
{\rm A}_1, {\rm B}_1, {\rm C}_1 \\
\mbox{\rm are imaginary.}
\end{array}
\]

The ${\rm A}_n$, ${\rm B}_n$, ${\rm C}_n$ being real or imaginary according to whether $n$ is even or odd, we can write\footnote{The notations were $p = \varpi + i \varpi'$, ${\rm A} = {\rm M} + i {\rm M}'$, ${\rm B} = {\rm N} + i {\rm N}'$ and ${\rm C} = {\rm P} + i {\rm P}'$ in the original (\emph{Translator's note}).}
\[
{\rm A} = {\cal M} + i {\cal M}', \quad {\rm B} = {\cal N} + i {\cal N}', \quad {\rm C} = {\cal P} + i {\cal P}' \quad \mbox{\rm and} \quad p = \alpha + i \beta .
\]
If we only consider average values of the displacements (that is, the constant parts of ${\rm A}$, etc.), we can put them in the form
\[
 u_0 + i u'_0 \quad , \qquad v_0 + i v'_0 \quad , \qquad w_0 + i w'_0
\]
where the portions ($u_0, v_0, w_0$) are even polynomials in $\omega$ and ($u'_0, v'_0, w'_0$) are odd ones. As far as the quantity $\frac{p}{\omega}$ is concerned, it must be an even function of $\omega$ and purely imaginary, since the governing equations contain only $\omega^2$. Indeed, I show directly in Appendix 2 that ${\rm P}_3$ and thus $p_2$ are zero in general.\footnote{Except for special cases such as the one to be studied below (\emph{Translator's note}).}

The values $u$, $v$ and $w$ thus take the form
\begin{eqnarray*}
 u & = & \left( u_0 + i u'_0 \right) e^{i \beta z - i \omega t} \quad , \\
 v & = & \left( v_0 + i v'_0 \right) e^{i \beta z - i \omega t} \quad , \\
 w & = & \left( w_0 + i w'_0 \right) e^{i \beta z - i \omega t} \quad ,
\end{eqnarray*}
where $\beta$ can take either of two values (Appendix 1), each of which corresponds to a set of values for $u$, $v$, etc.

The real parts of $u$, $v$ and $w$ must also satisfy the proposed equations because their coefficients are real; thus
\begin{eqnarray*}
 u & = & u_0 \cos \left( \beta z - \omega t \right) - u'_0 \sin \left( \beta z - \omega t \right) \quad , \qquad (*) \\
 v & = & v_0 \cos \left( \beta z - \omega t \right) - v'_0 \sin \left( \beta z - \omega t \right) \quad , \qquad (*) \\
 w & = & w_0 \cos \left( \beta z - \omega t \right) - w'_0 \sin \left( \beta z - \omega t \right) \quad ; \qquad (*)
\end{eqnarray*}
which represents an elliptical vibration propagating with a velocity $\frac{\omega}{\beta}$. And because $\frac{\beta}{\omega} = i \left( p_1 + p_3 \omega^2 + \cdots \right)$, where $p_1$ can take one of two distinct values,\footnote{assumed here to be negative imaginary so that $\beta/\omega > 0$ (\emph{Translator's note})} we see that two vibrations whose ellipticity depends on the wavelength can propagate as plane waves transverse to $z$ with two different velocities, and that the first term of the difference between these velocities is independent of the wavelength, in agreement with experiment. If we keep only the first terms in all the series, and thus set $u'_0$, $v'_0$, $w'_0$, $p_3$, etc. equal to zero, the double refraction remains but dispersion disappears.

From the results of Appendix 1, we could by changes of coordinates deduce the velocities corresponding to arbitrary directions of wave propagation. But if we want to limit ourselves to the first terms, it would be better to substitute the general values of $u$, $v$ and $w$ from (T2) directly into the differential equations expanded in powers of $\omega$, and by following a method identical to that followed above, the amplitude coefficients $a_0$, $b_0$ and $c_0$ are found to be determined by equations of the form
\begin{eqnarray*}
 \lambda^2 a_0 -m \left( m a_0 + n b_0 + p c_0 \right) & = & D a_0 + K b_0 + H c_0 \\
 \lambda^2 b_0 -n \left( m a_0 + n b_0 + p c_0 \right) & = & K a_0 + E b_0 + G c_0 \\
 \lambda^2 c_0 -p \left( m a_0 + n b_0 + p c_0 \right) & = & H a_0 + G b_0 + F c_0
\end{eqnarray*}
\[
\mbox{\rm where} \quad \lambda^2 = m^2 + n^2 + p^2
\]
which simultaneously determine the two velocities with which plane wave vibrations can be propagated. It is easily shown that these equations lead to a Fresnel wave surface, and produce a vibration transverse to the ray in the plane that projects it onto the plane of the wave. This form being less convenient than the one originally obtained for the special case of propagation only in the $z$-direction, I will revert to the latter in what follows.

The results above can be regarded as established only insofar as the ratios $a_0 : b_0 : c_0$ are really determined by equations (A3) of Appendix 1. However, the orientation of the plane of the wave relative to the medium might be such that these three equations reduce to two, one determining ${\rm P}_2$, the other only fixing the polarization of the wave (this is the general case for cubic media, the case of a wave bitangential to the wave surface by the bi-axes, or of a wave perpendicular to the axis). If we designate by ${\rm M}$, ${\rm M}'$, ${\rm M}''$, ${\rm N}$, ${\rm N}'$ and ${\rm N}''$ the six integrals\footnote{Explicitly,
\[ {\rm M} = \left[ \rho \left( \displaystyle\frac{\partial \varphi'}{\partial x} + 1 \right) \right] , \quad {\rm M}' = \left[ \rho \left( \displaystyle\frac{\partial \varphi''}{\partial y} + 1 \right) \right] , \quad {\rm M}'' = \left[ \rho \left( \displaystyle\frac{\partial \varphi'''}{\partial z} + 1 \right) \right]
\]
\[
{\rm N} =  \left[ \rho \displaystyle\frac{\partial \varphi'''}{\partial y} \right] = \left[ \rho \displaystyle\frac{\partial \varphi''}{\partial z} \right] , \quad {\rm N}' = \left[ \rho \displaystyle\frac{\partial \varphi'''}{\partial x} \right] = \left[ \rho \displaystyle\frac{\partial \varphi'}{\partial z} \right] , \quad {\rm N}'' = \left[ \rho \displaystyle\frac{\partial \varphi''}{\partial x} \right] = \left[ \rho \displaystyle\frac{\partial \varphi'}{\partial y} \right]
\]
(see Appendix 1 )\emph{(Translator's note)}.} that appear in these equations, they take the form
\begin{eqnarray*}
 \left( {\rm P}_2 + {\rm M} \right) a_0 + {\rm N}'' b_0 + {\rm N}' c_ 0 & = & 0, \\
 {\rm N}'' a_0 + \left( {\rm P}_2 + {\rm M}' \right) b_0 + {\rm N} c_0 & = & 0, \\
 {\rm N}' a_0 + {\rm N} b_0 + {\rm M}'' c_0 & = & 0 . \ (*)
\end{eqnarray*}

If we have ${\rm M}'' {\rm N}'' = {\rm N} {\rm N}'$ and ${\rm M} {\rm M}'' - {\rm N}^{\prime 2} = {\rm M}' {\rm M}'' - {\rm N}^{2} = - {\rm M}'' {\rm P}_2$, the values of ${\rm P}_2$ will be equal, and there will remain only the one equation
\[
{\rm N}' a_0 + {\rm N} b_0 + {\rm M}'' c_0 = 0
\]
to determine the ratios $a_0 : b_0 : c_0$; they are thus indeterminate.
So if we want to determine $a_1, b_1, c_1$ using equations (A4) from Appendix 2, under these conditions (A4) become
\[
{\rm M}'' {\rm P}_3 a_0 + {\rm N}' ({\rm N}' a_1 + {\rm N} b_1 + {\rm M}'' c_1) + p_1 {\rm M}'' \left[ \rho \frac{\partial \psi}{\partial x} \right] = 0,
\]
\[
{\rm M}'' {\rm P}_3 b_0 + {\rm N} ({\rm N}' a_1 + {\rm N} b_1 + {\rm M}'' c_1) + p_1 {\rm M}'' \left[ \rho \frac{\partial \psi}{\partial y} \right] = 0,
\]
\[
{\rm N}' a_1 + {\rm N} b_1 + {\rm M}'' c_1 + \left[ \rho \frac{\partial \psi}{\partial z} \right] p_1 + p_1 \left[ \rho \varphi \right] = 0 ;
\]
Multiplying these equations by $a_0$, $b_0$ and ${\rm M}'' c_0$ respectively, then adding the results gives
\[
{\rm M}'' {\rm P}_3 ( a_0^2 + b_0^2 ) + p_1 \left[ a_0 \rho \frac{\partial \psi}{\partial x} + b_0 \rho \frac{\partial \psi}{\partial y} + c_0 \rho \frac{\partial \psi}{\partial z} + \rho \varphi \right] {\rm M}'' = 0 .
\]

The term inside the brackets is always zero (Appendix 2), but since the ratio $a_0 : b_0$ is not necessarily real, we cannot further put ${\rm P}_3 = 0$. On the contrary, solving the equations (Appendix 4)
\[
\frac{{\rm M}'' \left( {\rm P}_3 a_0 + p_1 \left[ \rho \displaystyle\frac{\partial \psi}{\partial x} \right] \right)}{{\rm N}'} = \frac{{\rm M}'' \left( {\rm P}_3 b_0 + p_1 \left[ \rho \displaystyle\frac{\partial \psi}{\partial y} \right] \right)}{{\rm N}'} = p_1 \left[ \rho \frac{\partial \psi}{\partial z} \right] + p_1 \left[ \rho \varphi \right] \eqno{(7)}
\]
we find that $\displaystyle\frac{a_0}{b_0} = \pm i$, and that ${\rm P}_3$ is real; consequently $p_2$ is an imaginary quantity whose sign changes with that of $\displaystyle\frac{a_0}{b_0}$. For this particular orientation, the wave can therefore propagate as two vibrations whose projection onto the plane of the wave is circular, because $(a_0^2 + b_0^2) = 0$, and with different velocities, the plane of vibration being ${\rm N}' x + {\rm N} y + {\rm M}' z = 0$ since we are not looking for an approximation of any higher order than this.

For $p$ we obtain the two values $\omega p_1 \pm \omega^2 p_2$ $(*)$, both imaginary; if $\beta_1$ and $\beta_2$ are the imaginary parts of $p_1$ and $p_2$,\footnote{$\beta_{1,2}$ are $\varpi_{1,2}$ in the original (\emph{Translator's note}).} the values of the displacements being
\[
u = a_0 e^{i[(\beta_1 \omega \pm \beta_2 \omega^2)z - \omega t]}
\eqno{(*)}
\]
\[
v = \pm i a_0 e^{i[(\beta_1 \omega \pm \beta_2 \omega^2)z - \omega t]}
\eqno{(*)}
\]
and their real parts being
\[
u = a_0 \cos [(\beta_1 \omega \pm \beta_2 \omega^2)z - \omega t]
\eqno{(*)}
\]
\[
v =\mp  a_0 \sin [(\beta_1 \omega \pm \beta_2 \omega^2)z - \omega t],
\eqno{(*)}
\]
these are the equations of two circular vibrations propagating with the velocities $\displaystyle\frac{1}{\beta_1 \pm \beta_2 \omega}$ and wavelengths $\displaystyle\frac{2\pi}{\beta_1 \omega \pm \beta_2 \omega^2}$ $(*)$. Since the rotatory power is given by the difference between the inverses of the wavelengths of the two rays, it will be $\beta_2 \omega^2/\pi$ $(*)$; thus it will vary as the inverse square of the wavelength since $\omega = 2\pi/\tau$.

When the medium possesses a certain symmetry (if it is not merohedral), the values of $p_2$ and ${\rm P}_3$ are zero, and the ratio $a_0 : b_0$ is really indeterminate, which is most often the case.

There is no interest in pursuing this discussion by the examination of the terms that depend on higher powers of $\omega$ or of the inverse of the wavelength; the main goal I had set myself seems to have been achieved, since it follows from the calculations above that, under the sole condition that the expansions in ascending powers of the wavelength\footnote{Potier undoubtedly meant increasingly negative powers of the wavelength here (\emph{Translator's note}).} are admissible (and all analysts must make this assumption), equations with periodic coefficients account for the phenomena exhibited by transparent crystalline media.

\vspace*{.1in}
\hfill \rule{2in}{1pt} \hfill \hfill

\clearpage

\part*{APPENDICES}

\section*{Appendix 1}
Suppose a periodic function\footnote{In what follows, periodic functions except for $\rho$ are supposed to have had their constant parts (which do not contribute to their derivatives) removed. \emph{Translator's note:} In other words, all periodic functions except for $\rho$ are assumed to have an average value of $0$ over the elementary volume.} satisfies the differential equation
\[
\frac{\partial}{\partial x} \rho \frac{\partial F}{\partial x} + \frac{\partial}{\partial y} \rho \frac{\partial F}{\partial y} + \frac{\partial}{\partial z} \rho  \frac{\partial F}{\partial z} + U = 0 \eqno{({\rm A1})}
\]
($U$ being a function periodic like $\rho$); if $V$ is another periodic function, $d\varpi$ is a volume element, and the integral $\int$ is carried out over the elementary volume determined by the periodicity, we will have
\[
\int UV \, d\varpi = \int \rho \left( \frac{\partial F}{\partial x} \frac{\partial V}{\partial x} + \frac{\partial F}{\partial y} \frac{\partial V}{\partial y} + \frac{\partial F}{\partial z} \frac{\partial V}{\partial z} \right) \, d\varpi \eqno{({\rm A2})}
\]

Thus:
\begin{itemize}
\item[1.] If $U=0$ and $V=F$, then
\[
 \int \rho \left[ \left( \frac{\partial F}{\partial x} \right)^2 + \left( \frac{\partial F}{\partial y} \right)^2 + \left( \frac{\partial F}{\partial z} \right)^2 \right] \, d\varpi = 0
\]
and as a consequence the periodic function that satisfies the equation
\[
\frac{\partial}{\partial x} \rho \frac{\partial F}{\partial x} + \frac{\partial}{\partial y} \rho \frac{\partial F}{\partial y} + \frac{\partial}{\partial z} \rho  \frac{\partial F}{\partial x}  = 0 
\]
must be identically zero, and thus equation (A1) admits only one periodic solution.\footnote{Evidently, Potier assumes here that $\rho > 0$ everywhere (\emph{Translator's note}).} If there were two such solutions, their difference would satisfy the last equation above and would have to be zero.
\item[2.] Equation (4) therefore completely determines\footnote{The constants $a_0$, $b_0$ and $c_0$ being given (\emph{Translator's note}).} the function $\varphi$, and it is evident that if we put
\[
\frac{\partial}{\partial x} \rho \frac{\partial \varphi'}{\partial x} + \frac{\partial}{\partial y} \rho \frac{\partial \varphi'}{\partial y} + \frac{\partial}{\partial z} \rho  \frac{\partial \varphi'}{\partial x} = - \frac{\partial \rho}{\partial x}
\]
\[
\frac{\partial}{\partial x} \rho \frac{\partial \varphi''}{\partial x} + \frac{\partial}{\partial y} \rho \frac{\partial \varphi''}{\partial y} + \frac{\partial}{\partial z} \rho  \frac{\partial \varphi''}{\partial x} = - \frac{\partial \rho}{\partial y}
\]
\[
\frac{\partial}{\partial x} \rho \frac{\partial \varphi'''}{\partial x} + \frac{\partial}{\partial y} \rho \frac{\partial \varphi'''}{\partial y} + \frac{\partial}{\partial z} \rho  \frac{\partial \varphi'''}{\partial x} = - \frac{\partial \rho}{\partial z}
\]
then we will have $\varphi = a_0 \varphi' + b_0 \varphi'' + c_0 \varphi'''$.
\end{itemize}

Equations (5) can then be written
\[
\left.
\begin{array}{c}
a_0 {\rm P}_2 + \left[ \rho \left( \displaystyle\frac{\partial \varphi'}{\partial x} + 1 \right) \right] a_0 + \left[ \rho \displaystyle\frac{\partial \varphi''}{\partial x} \right] b_0 + \left[ \rho \displaystyle\frac{\partial \varphi'''}{\partial x} \right] c_0 = 0, \\
\\
\left[ \rho \displaystyle\frac{\partial \varphi'}{\partial y}  \right] a_0 + b_0 {\rm P}_2 + \left[ \rho \left( \displaystyle\frac{\partial \varphi''}{\partial y} + 1 \right) \right] b_0 + \left[ \rho \displaystyle\frac{\partial \varphi'''}{\partial y} \right] c_0 = 0, \ (*) \\
\\
\left[ \rho \displaystyle\frac{\partial \varphi'}{\partial z} \right] a_0 + \left[ \rho \displaystyle\frac{\partial \varphi''}{\partial z} \right] b_0 + \left[ \rho \left( \displaystyle\frac{\partial \varphi'''}{\partial z} + 1 \right) \right] c_0 = 0, \ (*)
\end{array} \right\} \eqno{({\rm A3})}
\]
which, after eliminating $a_0 : b_0 : c_0$, give a quadratic equation whose roots are the values of ${\rm P}_2$.\footnote{That is, setting the determinant of the system of equations (A3) equal to zero (\emph{Translator's note}).}

Of the nine integrals that appear in these equations, only six are independent. In fact, if in equation (A2) we first put $F = \varphi'$, $U = \frac{\partial \rho}{\partial x}$, $V = \varphi''$, then $F = \varphi''$, $U = \frac{\partial \rho}{\partial y}$, $V = \varphi'$, we get\footnote{Here and below I have added the notations ${\rm M}, \ldots , {\rm N}''$ used in the main body of the paper to designate the various integrals (\emph{Translator's note}).}
\[
\int \, d\varpi \, \rho \left( \frac{\partial \varphi'}{\partial x} \frac{\partial \varphi''}{\partial x} + \frac{\partial \varphi'}{\partial y} \frac{\partial \varphi''}{\partial y} + \frac{\partial \varphi'}{\partial z} \frac{\partial \varphi''}{\partial z} \right) = \int \varphi'' \frac{\partial \rho}{\partial x} \, d\varpi = \int \varphi' \frac{\partial \rho}{\partial y} \, d\varpi \equiv \frac{1}{V_p}{\rm N}''.
\]
where $V_p$ is the volume of the elementary parallelepiped. But the integrals 
\[
\int \, d\varpi \frac{\partial}{\partial x} \left( \rho \varphi'' \right) \quad \mbox{\rm and} \quad \int \, d\varpi \frac{\partial}{\partial y} \left( \rho \varphi' \right)
\]
evidently being zero, the two integrals appearing in the previous equation will be (apart from the signs)
\[
\int \rho \frac{\partial \varphi''}{\partial x}\, d\varpi \quad \mbox{\rm and} \quad \int \rho \frac{\partial \varphi'}{\partial y}\, d\varpi
\]

Likewise we can demonstrate the equalities
\[
\left[ \rho \frac{\partial \varphi'}{\partial z} \right] = \left[ \rho \frac{\partial \varphi'''}{\partial x} \right] \equiv {\rm N}' , \qquad \left[ \rho \frac{\partial \varphi''}{\partial z} \right] = \left[ \rho \frac{\partial \varphi'''}{\partial y} \right] \equiv {\rm N} .
\]
The left sides of equations (A3) are therefore the derivatives with respect to $a_0$, $b_0$ and $c_0$ of the same quadratic form; the ratios $a_0 : b_0 : c_0$ are thus real.\footnote{The quadratic form in question is
\[
Q = \frac{{\rm P}_2 + {\rm M}}{2} a_0^2 + \frac{{\rm P}_2 + {\rm M}'}{2} b_0^2 + \frac{{\rm M}''}{2} c_0^2 + {\rm N}'' a_0 b_0 + {\rm N}' a_0 c_0 + {\rm N} b_0 c_0
\]
(\emph{Translator's note}).}

Moreover, if we put $F = V = \varphi'$, $U = \frac{\partial \rho}{\partial x}$, equation (A2) becomes
\[
\int \, d\varpi \, \rho \left[ \left( \frac{\partial \varphi'}{\partial x} \right)^2 + \left( \frac{\partial \varphi'}{\partial y} \right)^2 + \left( \frac{\partial \varphi'}{\partial z} \right)^2 \right] = - \int \varphi' \frac{\partial \rho}{\partial x} \, d\varpi = + \int \rho \frac{\partial \varphi'}{\partial x} \, d\varpi
\]
and so the three integrals
\[
{\rm M} \equiv \left[ \rho \frac{\partial \varphi'}{\partial x} \right] , \quad {\rm M}' \equiv \left[ \rho \frac{\partial \varphi''}{\partial y} \right] , \quad {\rm M}'' \equiv \left[ \rho \frac{\partial \varphi'''}{\partial z} \right]
\]
are positive, and the values of ${\rm P}_2$ will both be real and negative.

\section*{Appendix 2}

Equation (T4) allows us to present an expression for $\psi$ composed of four parts.\footnote{The citation to this equation is corrected from the original (\emph{Translator's note}).} In fact, if we denote the unique periodic solution of the equation
\[
\frac{\partial}{\partial x} \rho \displaystyle\frac{\partial \Psi}{\partial x} + \displaystyle\frac{\partial}{\partial y} \rho \displaystyle\frac{\partial \Psi}{\partial y} + \frac{\partial}{\partial z} \rho \displaystyle\frac{\partial \Psi}{\partial z} + \rho \left( c_0 + \frac{\partial \varphi}{\partial z} \right) + \frac{\partial (\rho \varphi)}{\partial z} = 0
\]
by $\Psi$, we must have
\[
\psi = a_1 \varphi' + b_1 \varphi'' + c_1 \varphi''' + p_1 \Psi \eqno{(\rm T8)}
\]
so that equations (T5) become\footnote{The citation to this equation is corrected from the original (\emph{Translator's note}).}
\[
\left.
\begin{array}{c}
{\rm P}_3 a_0 + \left[ {\rm P}_2 + \rho \left( \displaystyle\frac{\partial \varphi'}{\partial x} + 1 \right) \right] a_1 + \left[ \rho \displaystyle\frac{\partial \varphi''}{\partial x} \right] b_1 + \left[ \rho \displaystyle\frac{\partial \varphi'''}{\partial x} \right] c_1 + p_1 \left[ \rho \displaystyle\frac{\partial \Psi}{\partial x} \right] = 0, \\
\\
{\rm P}_3 b_0 + \left[ \rho \displaystyle\frac{\partial \varphi'}{\partial y}  \right] a_1 +  \left[ {\rm P}_2 + \rho \left( \displaystyle\frac{\partial \varphi''}{\partial y} + 1 \right) \right] b_1 + \left[ \rho \displaystyle\frac{\partial \varphi'''}{\partial y} \right] c_1 + p_1 \left[ \rho \displaystyle\frac{\partial \Psi}{\partial y} \right] = 0, \\
\\
\left[ \rho \displaystyle\frac{\partial \varphi'}{\partial z} \right] a_1 + \left[ \rho \displaystyle\frac{\partial \varphi''}{\partial z} \right] b_1 + \left[ \rho \left( \displaystyle\frac{\partial \varphi'''}{\partial z} + 1 \right) \right] c_1 + p_1 \left[ \rho \varphi + \rho \displaystyle\frac{\partial \Psi}{\partial z} \right] = 0,
\end{array} \right\} \eqno{({\rm A4})}
\]

If we wish to use these equations to determine $a_1$, $b_1$, $c_1$, we observe that the matrix multiplying these unknowns is the same as that in equations (A3) that determine $a_0$, $b_0$, $c_0$; its determinant is thus zero. What is more, it is evident by comparing these two sets of equations that if $a_1$, $b_1$, $c_1$ satisfy equations (A4), then $a_1 + \lambda a_0$, $b_1 + \lambda b_0$, $c_1 + \lambda c_0$ also satisfy them; a term $\lambda \varphi$ will be added to the function $\psi$, and $\lambda {\rm A}_0$, $\lambda {\rm B}_0$, $\lambda {\rm C}_0$ to ${\rm A}_1$, ${\rm B}_1$, ${\rm C}_1$ respectively. Comparison of the equations that determine an arbitrary ${\rm A}_n$, ${\rm B}_n$, ${\rm C}_n$ with those that determine ${\rm A}_0$, ${\rm B}_0$, ${\rm C}_0$ shows that it must always turn out this way: the addition of $\lambda {\rm A}_0$, $\lambda {\rm B}_0$, $\lambda {\rm C}_0$ to ${\rm A}_n$, ${\rm B}_n$, ${\rm C}_n$ respectively amounts to multiplying $u$, $v$, $w$ by the factor $(1 + \lambda \omega^n)$.

The solvability condition for equations (A4) is obtained by multiplying them by $a_0$, $b_0$, $c_0$ respectively, adding the results and taking equations (A3) and the equalities demonstrated above for the six integrals into account; we get
\[
 {\rm P}_3 \left( a_0^2 + b_0^2 \right) + p_1 \left[ a_0 \rho \frac{\partial \Psi}{\partial x} + b_0 \rho \frac{\partial \Psi}{\partial y} + c_0 \rho \frac{\partial \Psi}{\partial z} \right] = 0 \eqno{({\rm A5})}
\]
Since ${\rm P}_3 = 2 p_1 p_2$, we see that $p_2$ will be real because $p_1$ is of the form $i{\rm S}$.

The first equation in this Appendix determines the function $\Psi$ $(*)$; since it is in the form of equation (A1) we deduce from it that
\[
 \int \rho \, d\varpi \left( \frac{\partial \Psi}{\partial x} \frac{\partial \varphi}{\partial x} + \frac{\partial \Psi}{\partial y} \frac{\partial \varphi}{\partial y} + \frac{\partial \Psi}{\partial z} \frac{\partial \varphi}{\partial z} \right) = c_0 \int \rho \varphi \, d\varpi
\]
On the other hand, putting
\[
 U = a_0 \frac{\partial \rho}{\partial x} + b_0 \frac{\partial \rho}{\partial y} + c_0 \frac{\partial \rho}{\partial z}, \qquad \mbox{\rm whence} \qquad F = \varphi \quad \mbox{\rm and} \quad V = \Psi,
\]
equation (A2) of Appendix 1 gives
\[
 \int \rho \, d\varpi \left( \frac{\partial \Psi}{\partial x} \frac{\partial \varphi}{\partial x} + \frac{\partial \Psi}{\partial y} \frac{\partial \varphi}{\partial y} + \frac{\partial \Psi}{\partial z} \frac{\partial \varphi}{\partial z} \right) = \int \, d\varpi \left( a_0 \frac{\partial \rho}{\partial x} + b_0 \frac{\partial \rho}{\partial y} + c_0 \frac{\partial \rho}{\partial z} \right) \Psi
\]
This last integral is equal to
\[
 - \int \, d\varpi \, \rho \left( a_0 \frac{\partial \Psi}{\partial x} + b_0 \frac{\partial \Psi}{\partial y} + c_0 \frac{\partial \Psi}{\partial z} \right) ;
\]
and thus, finally,
\[
 c_0 \int \rho \varphi \, d\varpi + \int \, d\varpi \, \rho \left( a_0 \frac{\partial \Psi}{\partial x} + b_0 \frac{\partial \Psi}{\partial y} + c_0 \frac{\partial \Psi}{\partial z} \right) = 0,
\]
and equation (A5) reduces to
\[
 {\rm P}_3 \left( a_0^2 + b_0^2 \right) = 0
\]
When $a_0 : b_0$ is real, ${\rm P}_2$ and $p_2$ are zero.

\section*{Appendix 3}
In this Appendix I obtain an expression for the quantity ${\rm P}_4 = p_2^2 + 2 p_1 p_3$ $(*)$ that determines the dispersion of the medium. We define a function $\Omega$ by the condition
\[
 \frac{\partial}{\partial x} \rho \frac{\partial \Omega}{\partial x} + \frac{\partial}{\partial y} \rho \frac{\partial \Omega}{\partial y} + \frac{\partial}{\partial z} \rho \frac{\partial \Omega}{\partial z} = \frac{\partial}{\partial x} \left( \rho A_2 \right) + \frac{\partial}{\partial y} \left( \rho B_2 \right) + \frac{\partial}{\partial z} \left( \rho C_2 \right)
\]
\[
  + p_1 \frac{\partial}{\partial z} \left( \rho \varphi_1 \right) + p_1^2 \frac{\partial}{\partial z} \left( \rho \Psi \right) + p_1^2 \rho \varphi + p_1 \left( \rho \frac{\partial \varphi_1}{\partial z} + c_1 \right),
\]
where $\varphi_1 = a_1 \varphi' + b_1 \varphi'' + c_1 \varphi'''$, and we end up with
\[
 {\rm P}_4 \left( a_0^2 + b_0^2 \right) + a_0 \left[ \rho \left( A_2 + \frac{\partial \Omega}{\partial x} \right) \right] + b_0 \left[ \rho \left( B_2 + \frac{\partial \Omega}{\partial y} \right) \right]
\]
It can be verified that the coefficient of each ${\rm P}_n$ will always be $\left( a_0^2 + b_0^2 \right)$.

\section*{Appendix 4}

We have shown in Appendix 2 that the solvability condition (A5) reduces to
\[
 {\rm P}_3 \left( a_0^2 + b_0^2 \right) = 0 .
\]
We need to show that in the case when the equation for ${\rm P}_2$ has two equal solutions, we must have
\[
 a_0^2 + b_0^2 = 0
\]
and that in general\footnote{Presumably Potier means here that ${\rm P}_3$ is real and not equal to zero; see below (\emph{Translator's note}).}
\[
 {\rm P}_3 \: \raisemath{-.7ex}{\stackrel{\textstyle <}{\textstyle >}} \: 0 .
\]
In fact, the equation that defines the function $\Psi$ shows that it is of the form
\[
 a_0 \Psi_1 + b_0 \Psi_2 + c_0 \Psi_3 ,
\]
if the $\Psi_n$ satisfy the equations
\[
 \frac{\partial}{\partial x} \rho \frac{\partial \Psi_1}{\partial x} + \frac{\partial}{\partial y} \rho \frac{\partial \Psi_1}{\partial y} + \frac{\partial}{\partial z} \rho \frac{\partial \Psi_1}{\partial z} + \frac{\partial}{\partial z} \left( \rho \varphi' \right) + \rho \frac{\partial \varphi'}{\partial z} = 0 ,
\]
\[
 \frac{\partial}{\partial x} \rho \frac{\partial \Psi_2}{\partial x} + \frac{\partial}{\partial y} \rho \frac{\partial \Psi_2}{\partial y} + \frac{\partial}{\partial z} \rho \frac{\partial \Psi_2}{\partial z} + \frac{\partial}{\partial z} \left( \rho \varphi'' \right) + \rho \frac{\partial \varphi''}{\partial z} = 0 ,
\]
\[
 \frac{\partial}{\partial x} \rho \frac{\partial \Psi_3}{\partial x} + \frac{\partial}{\partial y} \rho \frac{\partial \Psi_3}{\partial y} + \frac{\partial}{\partial z} \rho \frac{\partial \Psi_3}{\partial z} + \frac{\partial}{\partial z} \left( \rho \varphi''' \right) + \rho \frac{\partial \varphi'''}{\partial z} = 0 .
\]
The functions $\Psi_n$ defined in this way possess the following properties:
\[
 \left[ \rho \frac{\partial \Psi_1}{\partial x} \right] = \left[ \rho \frac{\partial \Psi_2}{\partial y} \right] = \left[ \rho \frac{\partial \Psi_1}{\partial y} + \rho \frac{\partial \Psi_2}{\partial x} \right] = 0 ,
\]
\[
 \left[ \rho \frac{\partial \Psi_1}{\partial z} + \rho \frac{\partial \Psi_3}{\partial x} + \rho \varphi' \right] = \left[ \rho \frac{\partial \Psi_2}{\partial z} + \rho \frac{\partial \Psi_3}{\partial y} + \rho \varphi'' \right] = \left[ \rho \frac{\partial \Psi_3}{\partial z} + \rho \varphi''' \right] = 0 ,
\]
which are proved in the same way as equation (A5) of Appendix 2.

Now multiplying both sides of the equation that determines $\Psi_2$ by $\varphi'''$ and both sides of the one that determines $\Psi_3$ by $\varphi''$, integrating over the elementary volume gives
\[
 \left[ \rho \left( \frac{\partial \Psi_2}{\partial x} \frac{\partial \varphi'''}{\partial x} + \frac{\partial \Psi_2}{\partial y} \frac{\partial \varphi'''}{\partial y} + \frac{\partial \Psi_2}{\partial z} \frac{\partial \varphi'''}{\partial z} \right) \right] \qquad \qquad \qquad \qquad \qquad \,
\]
\[
 = \int \, d\varpi \frac{\varphi'''}{\varphi''} \frac{\partial}{\partial z} \left( \rho \varphi^{\prime\prime 2} \right) = - \int \, d\varpi \rho \varphi^{\prime\prime 2} \frac{\partial}{\partial z} \frac{\varphi'''}{\varphi''} = \int \rho \, d\varpi \left( \varphi'' \frac{\partial \varphi'''}{\partial z} - \varphi''' \frac{\partial \varphi''}{\partial z} \right) \quad (*)
\]
and
\[
\left[ \rho \left( \frac{\partial \varphi''}{\partial x} \frac{\partial \Psi_3}{\partial x} + \frac{\partial \varphi''}{\partial y} \frac{\partial \Psi_3}{\partial y} + \frac{\partial \varphi''}{\partial z} \frac{\partial \Psi_3}{\partial z} \right) \right] \qquad \qquad \qquad \qquad \qquad \qquad \qquad \,
\]
\[
\qquad \qquad \quad \quad = - \int \rho \, d\varpi \left( \varphi''' \frac{\partial \varphi''}{\partial z} - \varphi'' \frac{\partial \varphi'''}{\partial z} \right) + \int \, d\varpi \, \rho \varphi''' . \quad (*)
\]
But the two integrals on the left sides are, thanks to equation (A2) of Appendix 1, equal to
\[
 - \int \, d\varpi \rho \frac{\partial \Psi_2}{\partial z} \qquad \mbox{\rm and} \qquad - \int \, d\varpi \rho \frac{\partial \Psi_3}{\partial z}
\]
respectively, if we put $F = \varphi'''$, $V = \Psi_2$, or $F = \varphi''$, $V = \Psi_3$. We thus have
\[
 \int \rho \frac{\partial \Psi_2}{\partial z} \, d\varpi = \int \, d\varpi \rho \left( \varphi'' \frac{\partial \varphi'''}{\partial z} - \varphi''' \frac{\partial \varphi''}{\partial z} \right) ,
\]
\[
 \int \rho \frac{\partial \Psi_3}{\partial y} \, d\varpi = \int \, d\varpi \rho \left( \varphi''' \frac{\partial \varphi''}{\partial z} - \varphi'' \frac{\partial \varphi'''}{\partial z} \right) - \int \, d\varpi \varphi''' ,
\]
and upon adding we get
\[
 \int \rho \left( \frac{\partial \Psi_2}{\partial z} + \frac{\partial \Psi_3}{\partial y} \right) \, d\varpi = - \int \, d\varpi \rho \varphi''' ,
\]
and likewise for the other relations to be proved.

Substituting the expressions deduced above for $\left[ \rho \displaystyle\frac{\partial \Psi}{\partial x} \right]$, $\left[ \rho \displaystyle\frac{\partial \Psi}{\partial y} \right]$ and $\left[ \rho \displaystyle\frac{\partial \Psi}{\partial z} \right]$ into equation (T5), and using (T8),\footnote{The citations to these equations are corrected from the original (\emph{Translator's note}).} they become
\[
 \left[ \rho \frac{\partial \Psi}{\partial x} \right] = b_0 \left[ \rho \frac{\partial \Psi_2}{\partial x} \right] + c_0 \left[ \rho \frac{\partial \Psi_3}{\partial x} \right] ,
\]
\[
 \left[ \rho \frac{\partial \Psi}{\partial y} \right] = a_0 \left[ \rho \frac{\partial \Psi_1}{\partial y} \right] + c_0 \left[ \rho \frac{\partial \Psi_3}{\partial y} \right] ,
\]
\[
 \left[ \rho \frac{\partial \Psi}{\partial z} \right] = a_0 \left[ \rho \frac{\partial \Psi_2}{\partial z} \right] + b_0 \left[ \rho \frac{\partial \Psi_2}{\partial z} \right] + c_0 \left[ \rho \varphi''' \right] .
\]
But $a_0$, $b_0$ and $c_0$ are related by
\[
 {\rm N}' a_0 + N b_0 + {\rm M}'' c_0 = 0 ,
\]
by which $c_0$ can be eliminated to get
\[
 {\rm M}'' \left[ \rho \frac{\partial \Psi}{\partial x} \right] = -{\rm N}' a_0 \left[ \rho \frac{\partial \Psi_3}{\partial x} \right] + b_0 \left\{ {\rm M}'' \left[ \rho \frac{\partial \Psi_2}{\partial x} \right] - N \left[ \rho \frac{\partial \Psi_3}{\partial x} \right] \right\} ,
\]
\[
 {\rm M}'' \left[ \rho \frac{\partial \Psi}{\partial y} \right] = a_0 \left[ {\rm M}'' \rho \frac{\partial \Psi_1}{\partial y} - {\rm N}' \rho \frac{\partial \Psi_3}{\partial y} \right] - N b_0 \left[ \rho \frac{\partial \Psi_3}{\partial y} \right] ,
\]
\[
 {\rm M}'' \left[ \rho \frac{\partial \Psi}{\partial z} \right] = a_0 \left[ {\rm M}'' \rho \frac{\partial \Psi_1}{\partial z} + {\rm N}' \rho \varphi''' \right] + b_0 \left[ {\rm M}'' \rho \frac{\partial \Psi_2}{\partial z} + N \rho \varphi''' \right] .
\]
Equation $(7)$:
\[
 {\rm M}'' {\rm P}_3 a_0 = {\rm N}' p_1 \left[ \rho \left( \frac{\partial \Psi}{\partial z} + \varphi \right) \right] - {\rm M}'' \left[ \rho \frac{\partial \Psi}{\partial z} \right] p_1
\]
therefore reduces (taking into account the relations among the integrals that were established above) to
\[
 a_0 {\rm M}'' {\rm P}_3 + b_0 p_1 \left[ {\rm M}'' \rho \frac{\partial \Psi_2}{\partial x} - N \rho \frac{\partial \Psi_3}{\partial x} + {\rm N}' \rho \frac{\partial \Psi_3}{\partial y} \right] = 0 ;
\]
the other equation gives
\[
 b_0 {\rm M}'' {\rm P}_3 + a_0 p_1 \left[ {\rm M}'' \rho \frac{\partial \Psi_1}{\partial y} + N \rho \frac{\partial \Psi_3}{\partial x} - {\rm N}' \rho \frac{\partial \Psi_3}{\partial y} \right] = 0 ;
\]
from which we deduce
\[
 \frac{a_0}{b_0} + \frac{b_0}{a_0} = 0 \qquad \mbox{\rm or} \qquad a_0^2 + b_0^2 = 0
\]
and
\[
 M_2'' {\rm P}_2^3 = - p_1^2 \left[ {\rm M}'' \rho \frac{\partial \Psi_2}{\partial x} - N \rho \frac{\partial \Psi_3}{\partial x} + {\rm N}' \rho \frac{\partial \Psi_3}{\partial y} \right] .
\]

Since $p_1^2$ is negative, we see that ${\rm P}_3^2$ will be positive, and that ${\rm P}_3$ will be real and can take one of two equal but opposite values, its sign being linked to that of $\displaystyle\frac{a_0}{b_0}$. Moreover, ${\rm P}_3 = 2 p_1 p_2$ $(*)$; $p_2$ will thus be imaginary if $p_1$ is, and will change sign with $\displaystyle\frac{a_0}{b_0}$.

The quantity
\[
 \left[ {\rm M}'' \rho \frac{\partial \Psi_2}{\partial x} - N \rho \frac{\partial \Psi_3}{\partial x} + {\rm N}' \rho \frac{\partial \Psi_3}{\partial y} \right]
\]
thus indicates, through its sign and magnitude, the direction and the rotation of the plane of polarization.

If we consider a uniaxial medium, and waves polarized transverse to the axis, these formulas simplify; in that case we have ${\rm N} = {\rm N}' = 0$, ${\rm C}=0$ and there remains
\[
 {\rm P}_3^2 = - p_1^2 \left[ \rho \frac{\partial \Psi_2}{\partial x} \right]^2 ,
\]
whence
\[
 p_2 = \pm i \left[ \rho \frac{\partial \Psi_2}{\partial x} \right] \qquad \mbox{\rm or} \qquad \mp i \left[ \rho \frac{\partial \Psi_1}{\partial y} \right] ;
\]
these will remain the same for a medium with cubic symmetry as well.

\section*{Appendix 5}

If it is desired merely to account for the possible effects of periodicity, it is convenient to examine the case where $\rho$ reduces to the sum of three functions, of $x$, of $y$ and of $z$ respectively; the solutions of the equations then reduce to quadratures.

\end{document}